\documentclass[aos,MSNbibl,seceqn,dvips]{arximspdf}
\usepackage{graphicx}
\doi{10.1214/13-AOS1147} 
\volume{41}
\issue{4}
\pubyear{2013}
\firstpage{2197}
\lastpage{2217}

\makeatletter
\newtheorem{thmm}{Theorem}[section]
\newtheorem{prop}[thmm]{Proposition}
\newproclaim{defn}[thmm]{Definition}
\newproclaim{example}[thmm]{Example}
\newtheorem{lem}[thmm]{Lemma}
\newtheorem{cor}[thmm]{Corollary}
\newproclaim{rem}[thmm]{Remark}

\def\span{\operatorname{span}}
\def\Tr{\operatorname{Tr}}
\def\D{\mathcal{D}}
\def\S{\mathcal{S}}
\def\H{\mathcal{H}}
\def\F{\mathcal{F}}
\def\T{\mathcal{T}}
\def\R{\mathbb{R}}
\def\C{\mathbb{C}}
\def\i{\sqrt{-1}}
\def\e{{e}}
\def\L{\mathcal{L}}
\def\N{\mathbb{N}}
\def\Re{\operatorname{Re}}
\def\Im{\operatorname{Im}}
\def\supp{\operatorname{supp}}
\def\rank{\operatorname{rank}}
\newcommand{\eqref}[1]{(\ref{#1})}
\makeatother

\begin{document}
\begin{frontmatter}

\title{Quantum local asymptotic normality based on a~new quantum
likelihood ratio}
\runtitle{QLAN based on a new quantum likelihood ratio}

\begin{aug}
\author[a]{\fnms{Koichi} \snm{Yamagata}\corref{}\ead[label=e1]{k-yamagata@cr.math.sci.osaka-u.ac.jp}},
\author[a]{\fnms{Akio} \snm{Fujiwara}\ead[label=e2]{fujiwara@math.sci.osaka-u.ac.jp}}
\and
\author[b]{\fnms{Richard D.} \snm{Gill}\ead[label=e3]{gill@math.leidenuniv.nl}}
\runauthor{K. Yamagata, A. Fujiwara and R. D. Gill}
\affiliation{Osaka University, Osaka University and Leiden University}
\address[a]{K. Yamagata\\
A. Fujiwara\\
Department of Mathematics\\
Osaka University\\
1-1 Machikaneyama\\
Toyonaka, Osaka 560-0043\\
Japan\\
\printead{e1}\\
\phantom{E-mail:\ }\printead*{e2}}

\address[b]{R. D. Gill\\
Mathematical Institute\\
Leiden University\\
P.O. Box 9512\\
2300 RA Leiden\\
The Netherlands\\
\printead{e3}}
\end{aug}

\received{\smonth{10} \syear{2012}}
\revised{\smonth{5} \syear{2013}}

%
\begin{abstract}
We develop a theory of local asymptotic normality in the quantum domain
based on a novel quantum analogue of the log-likelihood ratio.
This formulation is applicable to any quantum statistical model
satisfying a mild smoothness condition.
As an application, we prove the asymptotic achievability of the Holevo
bound for the local shift parameter.
\end{abstract}

%

\end{frontmatter}

\section{Introduction}\label{sec1}

Suppose that one has $n$ copies of a quantum system each in the same
state depending on an unknown parameter $\theta$, and one wishes to
estimate $\theta$ by making some measurement on the $n$ systems together.
This yields data whose distribution depends on $\theta$ and on the
choice of the measurement.
Given the measurement, we therefore have a classical parametric
statistical model, though not necessarily an i.i.d. model, since we are
allowed to bring the $n$ systems together before measuring the
resulting joint system as one quantum object.
In that case the resulting data need not consist of (a function of) $n$
i.i.d. observations, and a key quantum feature is that we can generally
extract more information about $\theta$ using such ``collective'' or
``joint'' measurements than when we measure the systems separately.
What is the best we can do as $n\to\infty$, when we are allowed to
optimise both over the measurement and over the ensuing data processing?
The objective of this paper is to study this question by extending the
theory of local asymptotic normality (LAN), which is known to form an
important part of the classical asymptotic theory, to quantum
statistical models.

Let us recall the classical LAN theory first.
Given a statistical model
$\S= \{ p_{\theta}; \theta\in\Theta\}$
on a probability space
$(\Omega, \F, \mu)$
indexed by a parameter
$\theta$ that ranges over an open subset
$\Theta$ of
$\R^{d}$,
let us introduce a local parameter $h:=\sqrt{n}(\theta-\theta_0)$
around a fixed $\theta_0\in\Theta$.
If the parametrisation $\theta\mapsto p_\theta$ is sufficiently smooth,
it is known that the statistical properties of the model
$ \{ p^{\otimes n}_{\theta_0+h/\sqrt{n}}; h\in\R^d \}$
is similar to that of the Gaussian shift model $ \{ N(h,J_{\theta
_{0}}^{-1}); h\in\R^{d} \}$ for large $n$, where $p^{\otimes
n}_\theta$ is the $n$th i.i.d. extension of $p_\theta$, and
$J_{\theta
_0}$ is the Fisher information matrix of the model $p_\theta$ at
$\theta_0$.
This property is called the local asymptotic normality of the model $\S
$ \cite{Vaart}.

More generally, a sequence $ \{ p_{\theta}^{(n)}; \theta\in\Theta
\subset\R^{d} \}$ of statistical models on $(\Omega^{(n)}, \F
^{(n)}, \mu^{(n)})$ is called \emph{locally asymptotically normal}
(LAN) at $\theta_{0}\in\Theta$
if there exist a $d\times d$ positive matrix $J$ and random vectors
$\Delta^{(n)}=(\Delta_1^{(n)}, \ldots,\break \Delta_d^{(n)})$ such that
$\Delta
^{(n)}\stackrel{0}{\rightsquigarrow} N(0,J)$ and
\[
\log\frac{p_{\theta_{0}+h/\sqrt{n}}^{(n)}}{p_{\theta_{0}}^{(n)}} =h^{i}\Delta_{i}^{(n)}-
\frac{1}{2}h^{i}h^{j}J_{ij}+o_{p_{\theta_0}}(1)
\]
for all $h\in\R^{d}$.
Here the arrow $\stackrel{h}{\rightsquigarrow}$ stands for the
convergence in distribution
under $p_{\theta_{0}+h/\sqrt{n}}^{(n)}$, the remainder term
$o_{p_{\theta_0}}(1)$ converges in probability to zero under
$p_{\theta
_0}^{(n)}$, and Einstein's summation convention is used.
The above expansion is similar in form to the log-likelihood ratio of
the Gaussian shift model:
\[
\log\frac{dN(h,J^{-1})}{dN(0,J^{-1})} \bigl(X^1,\ldots, X^d
\bigr)=h^i \bigl(X^j J_{ij} \bigr) -
\frac{1}{2}h^ih^j J_{ij}.
\]
This is the underlying mechanism behind the statistical similarities
between models
$ \{ p_{\theta_{0}+h/\sqrt{n}}^{(n)}; h\in\R^{d} \}$
and
$ \{ N(h,J^{-1}); h\in\R^{d} \}$.

In order to put the similarities to practical use, one needs some
mathematical devices.
In general, a statistical theory comprises two parts.
One is to prove the existence of a statistic that possesses a certain
desired property (direct part), and
the other is to prove the nonexistence of a statistic that exceeds
that property (converse part).
In the problem of asymptotic efficiency, for example,
the converse part, the impossibility to do asymptotically better than
the best which can be done in the limit situation, is ensured by the
following proposition, which is usually referred to as ``Le Cam's third
lemma'' \cite{Vaart}.

%
\begin{prop}
\label{propclecam3}
Suppose $ \{ p_{\theta}^{(n)}; \theta\in\Theta\subset\R^{d}
\} $
is LAN at $\theta_{0}\in\Theta$, with $\Delta^{(n)}$ and $J$ being
as above,
and let $X^{(n)}=(X_1^{(n)},\ldots, X_r^{(n)})$ be a sequence of random vectors.
If the joint distribution of $X^{(n)}$ and $\Delta^{(n)}$ converges to
a Gaussian distribution, in that
\[
\pmatrix{ X^{(n)}
\cr
\Delta^{(n)} } \stackrel{0} {
\rightsquigarrow} N \biggl(\pmatrix{0
\cr
0 }, \pmatrix{ \Sigma& \tau
\cr
^{t}\tau& J } \biggr),
\]
then $X^{(n)}\stackrel{h}{\rightsquigarrow}N(\tau h,\Sigma)$ for all
$h\in\R^{d}$.
Here $ ^{t}\tau$ stands for the transpose of $\tau$.
\end{prop}

Now, it appears from this lemma that it already tells us something
about the direct problem.
In fact, by putting $X^{(n)j}:=\sum_{k=1}^{d} [J^{-1}
]^{jk}\Delta_{k}^{(n)}$, we have
\[
\pmatrix{X^{(n)}
\cr
\Delta^{(n)} }\stackrel{0} {
\rightsquigarrow}N\biggl (\pmatrix{0
\cr
0 },\pmatrix{J^{-1} & I
\cr
I &
J } \biggr),
\]
so that $X^{(n)} \stackrel{h}{\rightsquigarrow}N(h,J^{-1})$ follows
from Proposition~\ref
{propclecam3}.
This proves the existence of an asymptotically efficient estimator for
$h$. In the real world, however, we do not know $\theta_0$ (obviously).
Thus, the existence of an asymptotically optimal estimator for $h$ does
not translate into the existence of an asymptotically optimal estimator
of~$\theta$. In fact, the usual way that Le Cam's third lemma is used
in the subsequent analysis is in order to prove the so-called
representation theorem, \cite{Vaart}, Theorem~7.10. This theorem can be
used to tell us in several precise mathematical senses that no
estimator can asymptotically do better than what can be achieved in the
limiting Gaussian model.

For instance, Van der Vaart's version of the representation theorem
leads to the asymptotic minimax theorem, telling us that the worst
behaviour of an estimator as $\theta$ varies in a shrinking (1 over
root $n$) neighbourhood of $\theta_0$ cannot improve on what we expect
from the limiting problem. This theorem applies to \emph{all} possible
estimators, but only discusses their \emph{worst} behaviour in a
neighbourhood of $\theta$. Another option is to use the representation
theorem to derive the convolution theorem, which tells us that \emph
{regular} estimators (estimators whose asymptotic behaviour in a small
neighbourhood of $\theta$ is more or less stable as the parameter
varies) have a limiting distribution which in a very strong sense is
more disperse than the optimal limiting distribution which we expect
from the limiting statistical problem.

This paper addresses a quantum extension of LAN (abbreviated as QLAN).
As in the classical statistics, one of the important subjects of QLAN
is to show the existence of an estimator (direct part) that enjoys
certain desired properties.
Some earlier works of asymptotic quantum parameter estimation theory
revealed the asymptotic achievability of the Holevo bound, a quantum
extension of the Cram\'er--Rao type bound (see Section B.1 and B.2 in
\cite{YFGsupp}).
Using a group representation theoretical method, Hayashi and Matsumoto
\cite{HayashiMatsumoto} showed that the Holevo bound for the quantum
statistical model $\S(\C^2)=\{\rho_\theta; \theta\in\Theta
\subset\R^3\}
$ comprising the totality of density operators on the Hilbert space $\H
\simeq\C^2$ is asymptotically achievable at a given single point
$\theta
_{0}\in\Theta$.
Following their work, Gu\c{t}\u{a} and Kahn \cite
{GutaQLANfor2,GutaQLANforD} developed a theory of strong QLAN, and
proved that the Holevo bound is asymptotically uniformly achievable
around a given $\theta_{0}\in\Theta$ for the quantum statistical model
$\S(\C^D)=\{\rho_\theta; \theta\in\Theta\subset\R^{D^2-1}\}$
comprising the totality of density operators on the finite dimensional
Hilbert space $\H\simeq\C^D$.
They proved that
an i.i.d. model $ \{ \rho_{\theta_0+h/\sqrt{n}}^{\otimes n}; h\in
\R
^{D^2-1} \}$
and a certain
quantum Gaussian shift model can be translated by quantum channels
to each other asymptotically.
Although their result is powerful, their QLAN has several drawbacks.
First of all, their method works only for i.i.d. extension of the
totality $\S(\H)$ of the quantum states on the Hilbert space~$\H$, and
is not applicable to generic submodels of $\S(\H)$.
Moreover, it makes use of a special parametrisation $\theta$ of $\S
(\H
)$, in which the change of eigenvalues and eigenvectors are treated as
essential.
Furthermore, it does not work if the reference state $\rho_{\theta_0}$
has a multiplicity of eigenvalues.
Since these difficulties are inevitable in representation theoretical
approach advocated by Hayashi and Matsumoto \cite{HayashiMatsumoto},
Gu\c{t}\u{a} and Jen\c{c}ov\'a \cite{GutaQLANweak} also tried a
different approach to QLAN via the Connes cocycle derivative,
which was put forward in the literature as an appropriate quantum
analogue of the likelihood ratio.
However they did not formally establish an expansion which would be
directly analogous to the classical LAN.
In addition, their approach is limited to faithful state models.

The purpose of the present paper is to develop a theory of weak QLAN
based on a new quantum extension of the log-likelihood ratio.
This formulation is applicable to any quantum statistical model
satisfying a mild smoothness condition, and is free from artificial
setups such as the use of a special coordinate system and/or
nondegeneracy of eigenvalues of the reference state at which QLAN works.
We also prove asymptotic achievability of the Holevo bound for the
local shift parameter $h$ that belong to a dense subset of $\R^d$.

This paper is organised as follows.
The main results are summarised in Section~\ref{secmainResults}.
We first introduce a novel type of quantum log-likelihood ratio, and
define a quantum extension of local asymptotic normality in a quite
analogous way to the classical LAN.
We then explore some basic properties of QLAN, including a sufficient
condition for an i.i.d. model to be QLAN, and a quantum extension of Le
Cam's third lemma.
Section~\ref{secappQLAN} is devoted to application of QLAN, including
the asymptotic achievability of the Holevo bound and asymptotic
estimation theory for some typical qubit models.
Proofs of main results are deferred to Section A of supplementary
material \cite{YFGsupp}.
Furthermore, since we assume some basic knowledge of quantum estimation
theory throughout the paper,
we provide, for the reader's convenience, a brief exposition of quantum
estimation theory in Section~B of supplementary material \cite{YFGsupp},
including quantum logarithmic derivatives, the commutation operator and
the Holevo bound
(Section B.1),
estimation theory for quantum Gaussian shift models
(Section B.2),
and estimation theory for pure state models
(Section B.3).

It is also important to notice the limits of this work, which means
that there are many open problems left to study in the future. In the
classical case, the theory of LAN builds, of course, on the rich theory
of convergence in distribution, as studied in probability theory. In
the quantum case, there still does not exist a full parallel theory.
Some of the most useful lemmas in the classical theory\vadjust{\goodbreak} simply are not
true when translated in the quantum domain. For instance, in the
classical case, we know that if the sequence of random variables $X_n$
converges in distribution to a random variable $X$, and at the same
time the sequence $Y_n$ converges in probability to a constant $c$,
then this implies joint convergence in distribution of $(X_n,Y_n)$ to
the pair $(X,c)$. The obvious analogue of this in the quantum domain is
simply untrue. In fact, there is not even a general theory of
convergence in distribution at all: there is only a theory of
convergence in distribution toward quantum Gaussian limits.
Unfortunately, even in this special case the natural analogue of the
just mentioned result simply fails to be true.

Because of these obstructions we are not at present able to follow the
standard route from Le Cam's third lemma to the representation theorem,
and from there to asymptotic minimax or convolution theorems.

However we believe that the paper presents some notable steps in this
direction. Moreover, just as with Le Cam's third lemma, one is able to
use the lemma to construct what can be conjectured to be asymptotically
optimal measurement and estimation schemes. We make some more remarks
on these possibilities later in the paper.

\section{Main results}
\label{secmainResults}
\subsection{Quantum log-likelihood ratio}

In developing the theory of QLAN, it is crucial what quantity one
should adopt as the quantum counterpart of the likelihood ratio.
One may conceive of the Connes cocycle
\[
[D\sigma, D\rho]_t:=\sigma^{\sqrt{-1}t} \rho^{-\sqrt{-1}t}
\]
as \emph{the} proper counterpart
since it plays an essential role in discussing the sufficiency of a
subalgebra in quantum information theory \cite{Petz}.
Nevertheless, we shall take a different route to the theory of QLAN,
paying attention to the fact that a ``quantum exponential family''
\[
\rho_{\theta}= \e^{({1}/{2}) (\theta L -\psi(\theta)I)} \rho_0 \e^{({1}/{2}) (\theta L -\psi(\theta)I)}
\]
inherits nice properties of the classical exponential family \cite
{{AmariNagaoka,FujiwaraNagaoka1995}}.

%
\begin{defn}[(Quantum log-likelihood ratio)]
\label{defqlikelihoodRatio}
We say a pair of density operators $\rho$
and $\sigma$ on a finite dimensional Hilbert space $\H$ are \textit{mutually
absolutely continuous,} $\rho\sim\sigma$ in symbols, if there
exist a Hermitian operator $\L$ that satisfies
\[
\sigma=\e^{({1}/{2})\L}\rho\e^{({1}/{2})\L}.
\]
We shall call such a Hermitian operator $\L$ a \textit{quantum log-likelihood
ratio}. When the reference states $\rho$ and $\sigma$ need to be specified,
$\L$ shall be denoted by $\L(\sigma|\rho)$, so that
\[
\sigma=\e^{({1}/{2})\L(\sigma|\rho)}\rho\e^{({1}/{2})\L(\sigma|\rho)}.
\]
We use the convention that $\L(\rho|\rho)=0$.
\end{defn}\eject

%
\begin{example}
We say a state on $\H\simeq\C^d$ is \emph{faithful} if its density
operator is positive definite.
Any two faithful states are always mutually absolutely continuous,
and the corresponding quantum log-likelihood ratio is unique.
In fact,
given $\rho>0$ and $\sigma>0$, they are related as $\sigma=\e
^{({1}/{2})\L(\sigma|\rho)}\rho\e^{({1}/{2})\L(\sigma
|\rho)}$,
where
\[
\L(\sigma|\rho)=2\log \bigl(\sqrt{\rho^{-1}}\sqrt{\sqrt {\rho} \sigma
\sqrt{\rho}}\sqrt{\rho^{-1}} \bigr).
\]
Note that $\Tr\rho\e^{({1}/{2})\L(\sigma|\rho)}$ is identical
to the fidelity between $\rho$ and $\sigma$, and $\e^{
({1}/{2})\L(\sigma|\rho)}$
is nothing but the operator geometric mean
$\rho^{-1}\#\sigma$,
where
$A\#B:=A^{1/2} (A^{-1/2}BA^{-1/2} )^{1/2}A^{1/2}$ for positive
operators $A,B$ \cite{KuboAndo}.
Since $A\# B=B\# A$, the quantum log-likelihood ratio can also be
written as
\[
\L(\sigma|\rho)=2\log \bigl(\sqrt{\sigma} (\sqrt{\sqrt {\sigma}\rho\sqrt{
\sigma}} )^{-1}\sqrt{\sigma} \bigr).
\]
\end{example}

%
\begin{example}
Pure states $\rho=\vert\psi\rangle \langle\psi\vert $ and
$\sigma=\vert\xi
\rangle \langle\xi\vert $
are mutually absolutely continuous if and only if $\langle\xi|\psi
\rangle \neq0$.
In fact, the ``only if'' part is obvious.
For the ``if'' part,
consider $\L(\sigma|\rho):=2\log R$ where
\[
R:=I+\frac{1}{\vert\langle\xi|\psi\rangle \vert}\vert\xi \rangle \langle\xi \vert -\vert\psi\rangle
\langle\psi\vert.
\]
Now
\[
\e^{({1}/{2})\L(\sigma|\rho)}\vert\psi\rangle =R\vert\psi \rangle =\frac{\langle\xi|\psi\rangle }{\vert\langle\xi|\psi
\rangle \vert}\vert\xi
\rangle,
\]
showing that $\rho\sim\sigma$.
\end{example}

%
\begin{rem}
In general, density operators $\rho$ and $\sigma$ are mutually
absolutely continuous
if and only if
%
%
\begin{equation}
\label{eqabsCont} \sigma\!\!\downharpoonleft_{\supp\rho} >0 \quad\mbox {and} \quad\rank
\rho=\rank\sigma,
\end{equation}
where $\sigma\!\!\downharpoonleft_{\supp\rho}$ denotes
the ``excision'' of $\sigma$,
the operator on the subspace $\supp\rho:=(\ker\rho)^\perp$ of $\H$
defined by
\[
\sigma\!\!\downharpoonleft_{\supp\rho}:=\iota_\rho^* \sigma
\iota_\rho,
\]
where $\iota_\rho\dvtx  \supp\rho\hookrightarrow\H$ is the inclusion map.
In fact, the ``only if'' part is immediate.
To prove the ``if'' part, let $\rho$ and $\sigma$ be represented in the
form of block matrices
\[
\rho= \pmatrix{ \rho_0 & 0
\cr
0 & 0 }, \qquad \sigma= \pmatrix{
\sigma_0 & \alpha
\cr
\alpha^* & \beta}
\]
with $\rho_0>0$.
Since the first condition in (\ref{eqabsCont}) is equivalent to
$\sigma_0>0$,
the matrix $\sigma$ is further decomposed as
\[
\sigma= E^* \pmatrix{ \sigma_0 & 0
\cr
0 & \beta-\alpha^*
\sigma_0^{-1} \alpha} E,\qquad E:= \pmatrix{ I &
\sigma_0^{-1} \alpha
\cr
0& I },
\]
and the second condition in (\ref{eqabsCont}) turns out to be
equivalent to $\beta-\alpha^* \sigma_0^{-1} \alpha=0$.
Now let $\L(\sigma|\rho):=2\log R$, where
\[
R:= E^* \pmatrix{ \rho_0^{-1} \# \sigma_0 &
0
\cr
0 & \gamma} E
\]
with $\gamma$ being an arbitrary positive matrix.
Then a simple calculation shows that $\sigma=R \rho R$.

The above argument demonstrates that a quantum log-likelihood ratio, if
it exists, is not unique when the reference states are not faithful.
To be precise, the operator $\e^{({1}/{2})\L(\sigma|\rho)}$ is
determined up to an additive constant Hermitian operator $K$ satisfying
$\rho K=0$.
This fact also proves that the quantity
$\Tr\rho\e^{({1}/{2})\L(\sigma|\rho)}$
is well defined regardless of the uncertainty of $\L(\sigma|\rho
)$, and is identical to the fidelity.
\end{rem}

\subsection{Quantum central limit theorem}

In quantum mechanics, canonical observables are represented by the
following canonical commutation relations (CCR):
\[
[Q_i,P_j]=\sqrt{-1} \hbar\delta_{ij} I,\qquad
[Q_i, Q_j]=0,\qquad  [P_i,P_j]=0,
\]
where $\hbar$ is the Planck constant.
In what follows, we shall treat a slightly generalised form of the CCR:
\[
\frac{\i}{2}[X_i,X_j] = S_{ij} I \qquad (1\leq
i,j\leq d),
\]
where $S=[S_{ij}]$ is a $d\times d$ real skew-symmetric matrix.
The algebra generated by the observables $(X_{1}, \ldots, X_{d})$ is
denoted by $\operatorname{CCR}(S)$,
and $X:=(X_{1}, \ldots, X_{d})$ is called the basic canonical
observables of the algebra $\operatorname{CCR}(S)$.
(See \cite{Holevo,qclt,CCR1,CCR2} for a rigorous definition of the CCR
algebra.)

A state $\phi$ on the algebra $\operatorname{CCR}(S)$
is characterised by the \emph{characteristic function}
\[
\F_{\xi}\{\phi\}:=\phi \bigl(\e^{\i\xi^{i}X_{i}} \bigr),
\]
where $\xi=(\xi^{i})_{i=1}^{d}\in\R^{d}$ and Einstein's summation
convention is used. A state $\phi$ on $\operatorname{CCR}(S)$ is
called a \textit{quantum
Gaussian state}, denoted by $\phi\sim N(h,J)$, if the characteristic
function takes the form
\[
\F_{\xi}\{\phi\}=\e^{\i\xi^{i}h_{i}-({1}/{2})\xi^{i}\xi^{j}V_{ij}},
\]
where $h=(h_{i})_{i=1}^{d}\in\R^{d}$ and $V=(V_{ij})$ is a real
symmetric matrix such that the Hermitian matrix $J:=V+\sqrt{-1}S$
is positive semidefinite. When the canonical observables $X$ need
to be specified, we also use the notation $(X,\phi)\sim N(h,J)$.
(See \cite{GillGuta,GutaUsta,Holevo,GutaQLANforD} for more information
about quantum Gaussian states.)

We will discuss relationships between a quantum Gaussian state $\phi$
on a CCR and a state on another algebra.\vadjust{\goodbreak}
In such a case,
we need to use the \emph{quasi-characteristic function}
%
%
\begin{eqnarray}\label{eqquasiChara}
&&\phi \Biggl(\prod_{t=1}^{r}
\e^{\i\xi_{t}^{i}X_{i}} \Biggr)
\nonumber
\\[-8pt]
\\[-8pt]
\nonumber
&&\qquad=\exp \Biggl(\sum_{t=1}^{r}
\biggl(\sqrt{-1}\xi_{t}^{i}h_{i}-
\frac{1}{2}\xi_{t}^{i}\xi_{t}^{j}J_{ji}
\biggr)-\sum_{t=1}^{r}\sum
_{s=t+1}^{r}\xi_{t}^{i}\xi
_{s}^{j}J_{ji} \Biggr),
\end{eqnarray}
of a quantum Gaussian state, where $(X,\phi)\sim N(h,J)$
and $\{\xi_{t}\}_{t=1}^{r}$ is a finite subset of $\C^{d}$ \cite{qclt}.

Given a sequence $\H^{(n)}$, $n\in\N$, of finite dimensional Hilbert
spaces, let
$X^{(n)}=(X_{1}^{(n)}, \ldots, X_{d}^{(n)})$ and $\rho^{(n)}$ be a list
of observables and a density operator on each~$\H^{(n)}$.
We say the sequence $ (X^{(n)},\rho^{(n)} )$ \emph{converges
in law to a quantum Gaussian state} $N(h,J)$, denoted as $(X^{(n)},\rho
^{(n)})\mathop{\rightsquigarrow}_{q}N(h,J)$, if
\[
\lim_{n\rightarrow\infty}\Tr\rho^{(n)} \Biggl(\prod
_{t=1}^{r}\e^{\sqrt
{-1}\xi_{t}^{i}X_{i}^{(n)}} \Biggr)=\phi \Biggl(\prod
_{t=1}^{r}\e^{\sqrt
{-1}\xi_{t}^{i}X_{i}} \Biggr)
\]
for any finite subset $\{\xi_{t}\}_{t=1}^{r}$ of $\C^{d}$, where
$(X,\phi)\sim N(h,J)$.
Here we do not intend to introduce the notion of ``quantum convergence
in law'' in general.
We use this notion only for quantum Gaussian states in the sense of
convergence of quasi-characteristic function.

The following is a version of the quantum central limit theorem (see
\cite{qclt}, e.g.).

%
\begin{prop}[(Quantum central limit theorem)]
\label{propqclt}
Let $A_{i}$ $(1\leq i\leq d)$ and $\rho$ be
observables and a state on a finite dimensional Hilbert space $\H$
such that $\Tr\rho A_{i}=0$, and let
\[
X_{i}^{(n)}:=\frac{1}{\sqrt{n}}\sum
_{k=1}^{n}I^{\otimes(k-1)}\otimes A_{i}
\otimes I^{\otimes(n-k)}.
\]
Then $(X^{(n)},\rho^{\otimes n})\mathop{\rightsquigarrow
}_{q}N(0,J)$, where $J$ is the
Hermitian matrix whose $(i,j)$th entry is given by $J_{ij}=\Tr\rho A_{j}A_{i}$.
\end{prop}

For later convenience, we introduce the notion of an ``infinitesimal''
object relative to the convergence $(X^{(n)},\rho^{(n)})\mathop
{\rightsquigarrow}_{q}N(0,J)$
as follows.
Given a list $X^{(n)}=(X_{1}^{(n)}, \ldots, X_{d}^{(n)})$ of observables
and a state $\rho^{(n)}$ on each $\H^{(n)}$ that satisfy
$(X^{(n)},\rho
^{(n)})\mathop{\rightsquigarrow}_{q}N(0,J)\sim(X,\phi)$,
we say a sequence $R^{(n)}$ of observables, each being defined on $\H
^{(n)}$, is \textit{infinitesimal relative to the convergence}
$(X^{(n)},\rho^{(n)})\mathop{\rightsquigarrow}_{q}N(0,J)$ if it satisfies
%
%
\begin{equation}
\lim_{n\rightarrow\infty}\Tr\rho^{(n)} \Biggl(\prod
_{t=1}^{r}\e^{\sqrt
{-1} (\xi_{t}^{i}X_{i}^{(n)}+\eta_{t}R^{(n)} )} \Biggr)=\phi \Biggl(\prod
_{t=1}^{r}\e^{\sqrt{-1}\xi_{t}^{i}X_{i}} \Biggr)\label
{eqinfinitesimal}
\end{equation}
for any finite subset of $ \{ \xi_{t} \} _{t=1}^{r}$ of $\C
^{d}$ and any finite subset $ \{ \eta_{t} \} _{t=1}^{r}$ of $\C$.
This is equivalent to saying that
\[
\biggl(\pmatrix{X^{(n)}
\cr
R^{(n)} },\rho^{(n)}
\biggr) \mathop{\rightsquigarrow}_{q} N \biggl(\pmatrix{0
\cr
0 }, \pmatrix{ J
& 0
\cr
0& 0 } \biggr),
\]
and is much stronger a requirement than
\[
\bigl(R^{(n)},\rho^{(n)} \bigr)\mathop{\rightsquigarrow}_{q}N(0,0).
\]
An infinitesimal object $R^{(n)}$ relative to $(X^{(n)},\rho
^{(n)})\mathop{\rightsquigarrow}_{q}N(0,J)$ will be denoted as
$o(X^{(n)},\rho^{(n)})$.

The following is in essence a simple extension of Proposition~\ref
{propqclt}, but will turn out to be useful in applications.

%
\begin{lem}
\label{lemoclt}
In addition to assumptions of Proposition~\ref{propqclt},
let $P(n)$, $n\in\mathbb{N}$, be a sequence of observables on $\H$,
and let
\[
R^{(n)}:=\frac{1}{\sqrt{n}}\sum_{k=1}^{n}I^{\otimes(k-1)}
\otimes P(n)\otimes I^{\otimes(n-k)}.
\]
If $\lim_{n\rightarrow\infty}P(n)=0$ and $\lim_{n\rightarrow\infty
}\sqrt{n} \Tr\rho P(n)=0$,
then $R^{(n)}=o(X^{(n)},\rho^{\otimes n})$.
\end{lem}

This lemma gives a precise criterion for the convergence of
quasi-charac\-teristic function for quantum Gaussian states.

\subsection{Quantum local asymptotic normality}

We are now ready to extend the notion of local asymptotic normality
to the quantum domain.
%
%
\begin{defn}[(QLAN)]
\label{defQLAN}Given a sequence $\H^{(n)}$ of finite dimensional
Hilbert spaces, let $\S^{(n)}= \{ \rho_{\theta}^{(n)}; \theta\in
\Theta\subset\R^{d} \} $
be a quantum statistical model on $\H^{(n)}$, where $\rho_{\theta
}^{(n)}$ is a
parametric family of density operators and $\Theta$ is an open set.
We say $\S^{(n)}$ is \textit{quantum locally asymptotically normal
}(QLAN) at $\theta_{0}\in\Theta$ if the following conditions are
satisfied:
\end{defn}
\begin{longlist}[(iii)]
\item[(i)] for any $\theta\in\Theta$ and $n\in\N$, $\rho_{\theta
}^{(n)}$ is
mutually absolutely continuous to~$\rho_{\theta_{0}}^{(n)}$,
\item[(ii)] there exist a list $\Delta^{(n)}=(\Delta_{1}^{(n)}, \ldots,
\Delta
_{d}^{(n)})$
of observables on each $\H^{(n)}$ that satisfies
\[
\bigl(\Delta^{(n)},\rho_{\theta_{0}}^{(n)} \bigr)\mathop {
\rightsquigarrow}_{q}N(0,J),
\]
where $J$ is a $d\times d$ Hermitian positive semidefinite matrix
with $\Re J>0$,\eject
\item[(iii)] quantum log-likelihood ratio $\L_{h}^{(n)}:=\L(\rho
_{\theta
_{0}+h/\sqrt{n}}^{(n)}\vert\rho_{\theta_{0}}^{(n)})$ is expanded in
$h\in\R
^d$ as
%
%
\begin{equation}
\L_{h}^{(n)} =h^{i}\Delta_{i}^{(n)}-
\tfrac{1}{2} \bigl(J_{ij}h^{i}h^{j}
\bigr)I^{(n)}+o \bigl(\Delta^{(n)}, \rho_{\theta_{0}}^{(n)}
\bigr), \label{eqqlanexpand}
\end{equation}
where $I^{(n)}$ is the identity operator on
$\H^{(n)}$.
\end{longlist}

It is also possible to extend Le Cam's third lemma (Proposition~\ref
{propclecam3}) to the quantum domain.
To this end, however, we need a device to handle the infinitesimal
residual term in (\ref{eqqlanexpand})
in a more elaborate way.

%
\begin{defn}
\label{defQLANX}
Let $\S^{(n)}= \{ \rho_{\theta}^{(n)}; \theta\in\Theta\subset\R
^{d} \} $
be as in Definition~\ref{defQLAN},
and let $X^{(n)}=(X_{1}^{(n)}, \ldots, X_{r}^{(n)})$ be a list
of observables on $\H^{(n)}$.
We say the pair $(\S^{(n)}, X^{(n)})$ is \emph{jointly
QLAN} at $\theta_{0}\in\Theta$ if the following conditions are satisfied:
\end{defn}
\begin{longlist}[(iii)]
\item[(i)] for any $\theta\in\Theta$ and $n\in\N$, $\rho_{\theta
}^{(n)}$ is
mutually absolutely continuous to~$\rho_{\theta_{0}}^{(n)}$,
\item[(ii)] there exist a list $\Delta^{(n)}=(\Delta_{1}^{(n)}, \ldots,
\Delta
_{d}^{(n)})$
of observables on each $\H^{(n)}$ that satisfies
%
%
\begin{equation}
\biggl(\pmatrix{X^{(n)}
\cr
\Delta^{(n)} },
\rho_{\theta_{0}}^{(n)} \biggr)\mathop{\rightsquigarrow}_{q}N
\biggl(\pmatrix{0
\cr
0 },\pmatrix{\Sigma& \tau
\cr
\tau^{*} & J }
\biggr), \label{eqqcovergenceTogether}
\end{equation}
where $\Sigma$ and $J$ are Hermitian positive semidefinite matrices of
size $r\times r$ and $d\times d$, respectively,
with $\Re J>0$, and $\tau$ is a complex matrix of size $r\times d$.
\item[(iii)] quantum log-likelihood ratio $\L_{h}^{(n)}:=\L(\rho
_{\theta
_{0}+h/\sqrt{n}}^{(n)}\vert\rho_{\theta_{0}}^{(n)})$ is expanded in
$h\in\R
^d$ as
%
%
\begin{equation}
\L_{h}^{(n)} =h^{i}\Delta_{i}^{(n)}-
\frac{1}{2} \bigl(J_{ij}h^{i}h^{j}
\bigr)I^{(n)}+o \biggl(\pmatrix{X^{(n)}
\cr
\Delta^{(n)}
}, \rho_{\theta_{0}}^{(n)} \biggr). \label{eqqlecam3expand}
\end{equation}
\end{longlist}

With Definition~\ref{defQLANX}, we can state a quantum extension of Le
Cam's third lemma as follows.
%
%
\begin{thmm}
\label{thmqlecam3}
Let $\S^{(n)}$ and $X^{(n)}$ be as in Definition~\ref{defQLANX}.
If $(\rho_{\theta}^{(n)}, X^{(n)})$ is jointly QLAN at $\theta
_{0}\in
\Theta$, then
\[
\bigl(X^{(n)}, \rho_{\theta_{0}+h/\sqrt{n}}^{(n)} \bigr)\mathop {
\rightsquigarrow}_{q}N \bigl( (\Re\tau)h, \Sigma \bigr)
\]
for any $h\in\R^{d}$.
\end{thmm}

It should be emphasised that assumption \eqref{eqqlecam3expand}, which
was superfluous in classical theory, is in fact crucial in proving
Theorem~\ref{thmqlecam3}.

In applications, we often handle i.i.d. extensions. In classical statistics,
a sequence of i.i.d. extensions of a model is LAN if the log-likelihood
ratio is twice differentiable
\cite{Vaart}.
Quite analogously, we can prove,
with\vadjust{\goodbreak} the help of Lemma~\ref{lemoclt},
that a sequence of i.i.d. extensions of a quantum statistical model is
QLAN if the quantum log-likelihood ratio is twice differentiable.
%
%
\begin{thmm}
\label{thmQLANiid}
Let $ \{ \rho_{\theta}; \theta\in\Theta\subset\R^{d} \} $
be a quantum statistical model on a finite dimensional Hilbert space
$\H$ satisfying $\rho_{\theta}\sim\rho_{\theta_{0}}$ for all
$\theta\in
\Theta$, where
$\theta_{0}\in\Theta$ is an arbitrarily fixed point.
If $\L_{h}:=\L(\rho_{\theta_{0}+h}|\rho_{\theta_{0}})$
is differentiable around $h=0$ and twice differentiable at $h=0$,
then $ \{ \rho_{\theta}^{\otimes n}; \theta\in\Theta\subset\R
^{d} \} $
is QLAN at $\theta_{0}$:
that is, $\rho_{\theta}^{\otimes n}\sim\rho_{\theta_{0}}^{\otimes
n}$, and
\[
\Delta_{i}^{(n)}:=\frac{1}{\sqrt{n}}\sum
_{k=1}^{n}I^{\otimes
(k-1)}\otimes L_{i}
\otimes I^{\otimes(n-k)}
\]
and $J_{ij}:=\Tr\rho_{\theta_{0}}L_{j}L_{i}$, with $L_{i}$ being the
$i$th symmetric logarithmic derivative at $\theta_{0}\in\Theta$,
satisfy conditions \textup{(ii)} and \textup{(iii)} in Definition~\ref{defQLAN}.
\end{thmm}

By combining Theorem~\ref{thmQLANiid} with Theorem~\ref{thmqlecam3}
and Lemma~\ref{lemoclt}, we obtain the following.

%
\begin{cor}
\label{corqlecam3iid}
Let $ \{ \rho_{\theta}; \theta\in\Theta\subset\R^{d} \}$
be a quantum statistical model on $\H$ satisfying $\rho_{\theta}\sim
\rho
_{\theta_{0}}$ for all $\theta\in\Theta$, where
$\theta_{0}\in\Theta$ is an arbitrarily fixed point.
Further, let $\{B_{i}\}_{1\le i\le r}$ be observables on $\H$
satisfying $\Tr\rho_{\theta_{0}}B_{i}=0$ for $i=1,\ldots,r$.
If $\L_{h}:=\L(\rho_{\theta_{0}+h}|\rho_{\theta_{0}})$ is
differentiable
around $h=0$ and twice differentiable at $h=0$,
then the pair $ ( \{ \rho_{\theta}^{\otimes n} \},
X^{(n)} )$ of i.i.d. extension model $ \{\rho_{\theta
}^{\otimes n} \}$ and the list $X^{(n)}=\{X^{(n)}_i\}_{1\le i\le
r}$ of observables defined by
\[
X_{i}^{(n)}:=\frac{1}{\sqrt{n}}\sum
_{k=1}^{n}I^{\otimes(k-1)}\otimes B_{i}
\otimes I^{\otimes(n-k)}
\]
is jointly QLAN at $\theta_{0}$, and
\[
\bigl(X^{(n)},\rho_{\theta_{0}+h/\sqrt{n}}^{\otimes n} \bigr)\mathop{
\rightsquigarrow}_{q}N \bigl( (\Re\tau)h,\Sigma \bigr)
\]
for any $h\in\R^{d}$,
where $\Sigma$ is the $r\times r$ positive semidefinite matrix defined
by $\Sigma_{ij}=\Tr\rho_{\theta_{0}}B_{j}B_{i}$ and
$\tau$ is the $r\times d$ matrix defined by $\tau_{ij}=\Tr\rho
_{\theta
_0}L_j B_i$
with $L_i$ being the $i$th symmetric logarithmic derivative at $\theta_0$.
\end{cor}

Corollary~\ref{corqlecam3iid} is an i.i.d. version of the quantum Le
Cam third lemma, and will play a key role in demonstrating the
asymptotic achievability of the Holevo bound.

\section{Applications to quantum statistics}
\label{secappQLAN}

\subsection{Achievability of the Holevo bound}

Corollary~\ref{corqlecam3iid} prompts us
to expect that, for sufficiently large $n$, the estimation problem for
the parameter $h$ of $\rho_{\theta_{0}+h/\sqrt{n}}^{\otimes n}$ could
be reduced to that for the shift parameter $h$ of the quantum Gaussian
shift model $N( (\Re\tau)h, \Sigma)$.
The\vadjust{\goodbreak} latter problem has been well-established to date
(see Section B.2 in \cite{YFGsupp}).
In particular, the best strategy for estimating the shift parameter $h$
is the one that achieves the Holevo bound\vspace*{1pt} $C_{h} (N( (\Re\tau
)h,\Sigma), G )$ (see Theorem B.7 in \cite{YFGsupp}).
Moreover, it
is
shown (see Corollary B.6 in \cite{YFGsupp}) that
the Holevo bound $C_{h} (N( (\Re\tau)h,\Sigma), G )$
is identical to the Holevo bound $C_{\theta_{0}} (\rho_{\theta},
G )$ for the model $\rho_\theta$ at $\theta_0$.
These facts suggest the existence of a sequence $M^{(n)}$ of estimators
for the parameter $h$ of $ \{\rho_{\theta_{0}+h/\sqrt{n}}^{\otimes
n} \}_n$ that asymptotically achieves the Holevo bound $C_{\theta
_{0}} (\rho_{\theta}, G )$.
The following theorem materialises this program.

%
\begin{thmm}
\label{thmachieveHolevo}
Let $ \{ \rho_{\theta}; \theta\in\Theta\subset\R^{d} \} $
be a quantum statistical model on a finite dimensional Hilbert space
$\H$, and fix a point $\theta_{0}\in\Theta$. Suppose that
$\rho_{\theta}\sim\rho_{\theta_{0}}$ for all $\theta\in\Theta$,
and that the quantum log-likelihood ratio $\L_{h}:=\L(\rho
_{\theta
_{0}+h}|\rho_{\theta_{0}})$
is differentiable in $h$ around $h=0$ and twice differentiable at
$h=0$. For any countable dense subset $D$ of $\R^{d}$ and any weight
matrix $G$, there exist a sequence $M^{(n)}$ of estimators on the
model $ \{ \rho_{\theta_{0}+h/\sqrt{n}}^{\otimes n}; h\in\R
^{d} \} $
that enjoys
\[
\lim_{n\rightarrow\infty}E_{h}^{(n)}
\bigl[M^{(n)} \bigr]=h
\]
and
\[
\lim_{n\rightarrow\infty}\Tr GV_{h}^{(n)}
\bigl[M^{(n)} \bigr]=C_{\theta_{0}} (\rho_{\theta},G )
\]
for every $h\in D$. Here $C_{\theta_{0}} (\rho_{\theta},G )$
is the Holevo bound at $\theta_{0}$.
Here $E_{h}^{(n)}[ \cdot]$ and $V_{h}^{(n)}[ \cdot]$ stand for the
expectation and the covariance matrix under the state $ \rho_{\theta
_{0}+h/\sqrt{n}}^{\otimes n}$.
\end{thmm}

Theorem~\ref{thmachieveHolevo} asserts that there is a sequence
$M^{(n)}$ of estimators on\break $ \{\rho_{\theta_{0}+h/\sqrt
{n}}^{\otimes n} \}_n$
that is asymptotically unbiased and achieves the Holevo bound
$C_{\theta
_{0}} (\rho_{\theta},G )$
for all $h$ that belong to a dense subset of $\R^{d}$.
Since this result requires only twice differentiability of the quantum
log-likelihood
ratio of the base model $\rho_\theta$, it will be useful in a wide
range of statistical estimation problems.

\subsection{Application to qubit state estimation}

In order to demonstrate the
applicability of our theory, we explore qubit state estimation problems.

%
\begin{example}[(3-dimensional faithful state model)]\label{ex3d}
The first example is an ordinary one, comprising the totality of
faithful qubit states:
\[
\S \bigl(\C^2 \bigr)= \bigl\{ \rho_{\theta}=\tfrac{1}{2}
\bigl(I+\theta^{1}\sigma_{1}+\theta^{2}
\sigma_{2}+\theta^{3}\sigma_{3} \bigr); \theta=
\bigl(\theta^i \bigr)_{1\le i\le3} \in\Theta \bigr\},
\]
where
$\sigma_{i}$ ($i=1,2,3$) are the standard Pauli matrices and $\Theta$
is the open unit ball in~$\R^3$.
Due to the rotational symmetry, we take the reference point to be
$\theta_0=(0,0,r)$, with $0\le r<1$.
By a direct calculation, we see that the
symmetric logarithmic derivatives\vadjust{\goodbreak} (SLDs)
of the model $\rho_\theta$ at $\theta=\theta_{0}$
are $(L_1, L_2, L_3)= (\sigma_{1}, \sigma_{2}, (rI+\sigma_{3})^{-1}
)$,
and the SLD Fisher information matrix $J^{(S)}$ at $\theta_0$ is given
by the real part of the matrix
\[
J:= [\Tr\rho_{\theta_{0}}L_{j}L_{i} ]_{ij}=
\pmatrix{1 & -r\sqrt{-1} & 0
\vspace*{2pt}\cr
r\sqrt{-1} & 1 & 0
\vspace*{2pt}\cr
0 & 0 & 1/ \bigl(1-r^{2} \bigr) }.
\]

Given a $3\times3$ real positive definite matrix $G$, the minimal value
of the weighted covariances at $\theta=\theta_0$ is given by
\[
\min_{\hat{M}}\Tr GV_{\theta_{0}}[\hat{M}]=C_{\theta_{0}}^{(1)}
(\rho_{\theta},G ),
\]
where the minimum is taken over all estimators $\hat M$ that are
locally unbiased at $\theta_0$, and
\[
C_{\theta_{0}}^{(1)} (\rho_{\theta},G ) = \bigl( \Tr\sqrt{
\sqrt{G} J^{(S)^{-1}} \sqrt{G} } \bigr)^{2}
\]
is the Hayashi--Gill--Massar bound \cite{GillMassar,Hayashi} (see also
\cite{YamagataTomo}).
On the other hand,
the SLD tangent space (i.e., the linear span of the SLDs) is obviously
invariant under the action of the commutation operator $\D$,
and the Holevo bound is
given by
\[
C_{\theta_{0}} (\rho_{\theta},G ):=\Tr GJ^{(R)^{-1}}+\Tr \bigl\vert
\sqrt{G} \Im J^{(R)^{-1}}\sqrt{G} \bigr\vert,
\]
where
\[
J^{(R)^{-1}}:=(\Re J)^{-1}J(\Re J)^{-1}=\pmatrix{1 & -r
\sqrt{-1} & 0
\cr
r\sqrt{-1} & 1 & 0
\cr
0 & 0 & 1-r^{2} }
\]
is the inverse of the right logarithmic derivative (RLD) Fisher
information matrix
(see Corollary B.2 in \cite{YFGsupp}).

It can be shown that the Hayashi--Gill--Massar bound is greater than the
Holevo bound:
\[
C_{\theta_0}^{(1)} (\rho_{\theta},G )>C_{\theta_{0}} (\rho
_{\theta},G ).
\]
Let us check this fact for the special case when $G=J^{(S)}$.
A direct computation shows that
\[
C_{\theta_0}^{(1)} \bigl(\rho_{\theta},J^{(S)}
\bigr)=9
\]
and
\[
C_{\theta_0} \bigl(\rho_{\theta},J^{(S)} \bigr)=3+2r.
\]
The left panel of Figure~\ref{figbounds} shows the behaviour of
$C_{\theta_0} (\rho_{\theta},J^{(S)} )$
(solid) and
$C_{\theta_0}^{(1)} (\rho_{\theta},J^{(S)} )$ (dashed) as
functions of $r$.
We see that the Holevo bound $C_{\theta_0} (\rho_{\theta},J^{(S)}
)$
is much smaller than $C_{\theta_0}^{(1)} (\rho_{\theta},J^{(S)}
)$.

%
\begin{figure}

\includegraphics{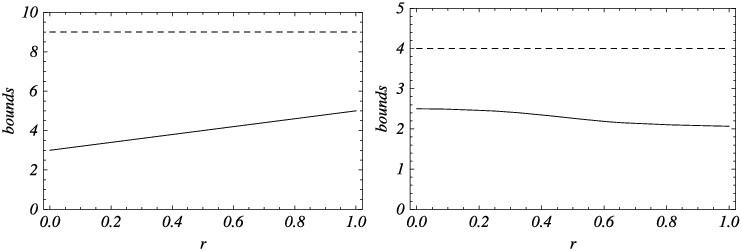}

\caption{The left panel displays the Holevo bound $C_{(0,0,r)} (\rho
_{\theta},J^{(S)} )$ (solid)
and the Hayashi--Gill--Massar bound $C_{(0,0,r)}^{(1)} (\rho_{\theta
},J^{(S)} )$ (dashed)
for the 3-D model $\rho_{\theta}=\frac{1}{2} (I+\theta^{1}\sigma
_{1}+\theta^{2}\sigma_{2}+\theta^{3}\sigma_{3} )$ as functions of
$r=\|\theta\|$.
The right panel displays the Holevo bound
$C_{(0,r)} (\rho_{\theta},J^{(S)} )$ (solid)
and the Nagaoka bound $C_{(0,r)}^{(1)} (\rho_{\theta},J^{(S)}
)$ (dashed)
for the 2-D model
$\rho_{\theta}=\frac{1}{2} (
I+\theta^{1}\sigma_{1}+\theta^{2}\sigma_{2}+\frac{1}{4}\sqrt{1-\|
\theta
\|^2} \sigma_3
)$.}
\label{figbounds}
\end{figure}

Does this fact imply that the Holevo bound is of no use? The answer is
contrary, as Theorem~\ref{thmachieveHolevo} asserts.
We will demonstrate the asymptotic achievability of the Holevo bound.
Let
\[
\Delta_{i}^{(n)}:=\frac{1}{\sqrt{n}}\sum
_{k=1}^{n}I^{\otimes
k-1}\otimes L_{i}
\otimes I^{\otimes n-k}
\]
and let $X_i^{(n)}:=\Delta_i^{(n)}$ for $i=1,2,3$.
It follows from the quantum central limit theorem that
\[
\biggl(\pmatrix{X^{(n)}
\cr
\Delta^{(n)} },
\rho_{\theta_{0}}^{\otimes n} \biggr)\mathop{\rightsquigarrow}_{q}N
\biggl(0,\pmatrix{J & J
\cr
J & J } \biggr).
\]
Since
\[
\L(\theta):=\L(\rho_{\theta}|\rho_{\theta_{0}})=2\log \Bigl(\sqrt
{\rho_{\theta_{0}}^{-1}}\sqrt{\sqrt{\rho_{\theta_{0}}}
\rho_{\theta}\sqrt{\rho_{\theta_{0}}}}\sqrt{
\rho_{\theta_{0}}^{-1}} \Bigr)
\]
is obviously of class $C^{\infty}$ in $\theta$,
Corollary~\ref{corqlecam3iid} shows that
$ ( \{ \rho_{\theta}^{\otimes n} \}, X^{(n)} )$
is jointly QLAN at $\theta_{0}$, and that
\[
\bigl(X^{(n)},\rho_{\theta_{0}+h/\sqrt{n}}^{\otimes n} \bigr)\mathop{
\rightsquigarrow}_{q}N \bigl(( \Re J)h,J \bigr)
\]
for all $h\in\R^3$.
This implies that a sequence of models $ \{\rho^{\otimes n}_{\theta
_0+h/\sqrt{n}}; h\in\R^d \}$ converges to a quantum Gaussian
shift model $ \{N((\Re J)h, J); h\in\R^3 \}$.
Note that the imaginary part
\[
S= \pmatrix{0 & -r\sqrt{-1} & 0
\cr
r\sqrt{-1} & 0 & 0
\cr
0 & 0 & 0 }
\]
of the matrix $J$ determines the $\operatorname{CCR}(S)$, as well as the
corresponding basic canonical observables $X=(X^1,X^2,X^3)$. When
$r\neq0$, the above $S$ has the following physical interpretation:
$X^1$ and $X^2$ form a canonical pair of quantum Gaussian observables,
while $X^3$ is a classical Gaussian random variable. In this way, the
matrix $J$ automatically tells us the structure of the limiting quantum
Gaussian shift model.

Now, the best strategy for estimating the shift parameter $h$ of the
quantum Gaussian shift model $ \{N((\Re J)h, J); h\in\R^d \}$
is the one that achieves the Holevo bound $C_{h} (N( (\Re
J )h, J), G )$
(see Theorem B.7 in \cite{YFGsupp}).
Moreover, this Holevo bound $C_{h} (N( (\Re J )h,J),
G )$ is identical to the Holevo bound $C_{\theta_{0}} (\rho
_{\theta}, G )$ for the model $\rho_\theta$ at $\theta_0$
(see Corollary B.6 in \cite{YFGsupp}.
Recall that the matrix $J$ is evaluated at $\theta_0$ of the model
$\rho
_\theta$).
Theorem~\ref{thmachieveHolevo} combines these facts, and concludes
that there exist a sequence $M^{(n)}$ of estimators on the model
$ \{ \rho_{\theta_{0}+h/\sqrt{n}}^{\otimes n}; h\in\R^{3} \} $
that is asymptotically unbiased and achieves the common values of the
Holevo bound:
\[
\lim_{n \rightarrow\infty} \Tr G V_{h}^{(n)}
\bigl[M^{(n)} \bigr] =C_{h} \bigl(N \bigl((\Re J)h,J \bigr),G
\bigr) =C_{\theta_0} (\rho_{\theta}, G )
\]
for all $h$ that belong to a countable dense subset of $\R^3$.

It should be emphasised that the matrix $J$ becomes the identity at the
origin $\theta_0=(0,0,0)$.
This means that the limiting Gaussian shift model $ \{N(h, J); h\in
\R^{3} \} $ is ``classical.''
Since such a degenerate case cannot be treated in \cite{GutaQLANfor2,HayashiMatsumoto,GutaQLANforD},
our method has a clear advantage in applications.
\end{example}

%
\begin{example}[(Pure state model)]
The second example is to demonstrate that our formulation allows us to
treat pure state models.
Let us consider the model
$\S=\{|\psi(\theta)\rangle\langle\psi(\theta)|; \theta=(\theta
^i)_{1\le i\le2} \in\Theta\}$ defined by
\[
\psi(\theta):=\frac{1}{\sqrt{\cosh\Vert\theta\Vert}} \e ^{({1}/{2}) (\theta^{1}\sigma_{1}+\theta^{2}\sigma_{2} )} \pmatrix{ 1
\cr
0 },
\]
where $\Theta$ is an open subset of $\R^2$ containing the origin, and
$\| \cdot\|$ denotes the Euclid norm.
By a direct computation, the SLDs at $\theta_0=(0,0)$ are $(L_1,
L_2)=(\sigma_{1}, \sigma_{2})$,
and the SLD Fisher information matrix $J^{(S)}$ is the real part of the matrix
\[
J= [\Tr\rho_{\theta_{0}}L_{j}L_{i} ]_{ij}=
\pmatrix{1 & -\sqrt{-1}
\cr
\sqrt{-1} & 1 },
\]
that is, $J^{(S)}=I$.
Since the SLD tangent space is $\D$ invariant \cite{fujiwaraCoherent},
the Holevo bound for a weight $G>0$ is represented as
\[
C_{\theta_0} (\rho_{\theta},G ):=\Tr GJ^{(R)^{-1}}+\Tr \bigl\vert
\sqrt{G} \Im J^{(R)^{-1}}\sqrt{G} \bigr\vert,
\]
where
\[
J^{(R)^{-1}}:=(\Re J)^{-1}J(\Re J)^{-1}=\pmatrix{1 & -
\sqrt{-1}
\cr
\sqrt{-1} & 1 }
\]
is the inverse RLD Fisher information matrix
(see Corollary B.2 in \cite{YFGsupp}).\vadjust{\goodbreak}

Let us demonstrate that our QLAN is applicable also to pure state models.
Let
\[
\Delta_{i}^{(n)}:=\frac{1}{\sqrt{n}}\sum
_{k=1}^{n}I^{\otimes
k-1}\otimes L_{i}
\otimes I^{\otimes n-k}
\]
and let $X_{i}^{(n)}:=\Delta_{i}^{(n)}$ for $i=1,2$.
It follows from the quantum central limit theorem that
\[
\biggl(\pmatrix{X^{(n)}
\cr
\Delta^{(n)} },
\rho_{\theta_{0}}^{\otimes n} \biggr)\mathop{\rightsquigarrow}_{q}N
\biggl(0,\pmatrix{J & J
\cr
J & J } \biggr).
\]
Since
\[
\L(\theta):=\L(\rho_{\theta}|\rho_{\theta_{0}})= \theta^{1}
\sigma_{1}+\theta^{2} \sigma_{2}-\log\cosh\Vert
\theta\Vert
\]
is of class $C^{\infty}$ with respect to $\theta$,
it follows from Corollary~\ref{corqlecam3iid} that $ ( \{ \rho
_{\theta}^{\otimes n} \},\break X^{(n)} )$
is jointly QLAN at $\theta_{0}$, and that
\[
\bigl(X^{(n)},\rho_{\theta_{0}+h/\sqrt{n}}^{\otimes n} \bigr)\rightsquigarrow
N \bigl((\Re J)h,J \bigr)=N \bigl(h,J^{(R)^{-1}} \bigr)
\]
for all $h\in\R^2$.
Theorem~\ref{thmachieveHolevo} further asserts that
there exist a sequence $M^{(n)}$ of estimators on the model
$ \{ \rho_{\theta_{0}+h/\sqrt{n}}^{\otimes n}; h\in\R^{2} \} $
that is asymptotically unbiased and achieves the Holevo bound:
\[
\lim_{n \rightarrow\infty} \Tr G V_{h}^{(n)}
\bigl[M^{(n)} \bigr] =C_{h} \bigl(N \bigl(h,J^{(R)^{-1}}
\bigr), G \bigr)=C_{(0,0)} (\rho_{\theta}, G )
\]
for all $h$ that belong to a dense subset of $\R^3$.
In fact, the sequence $M^{(n)}$ can be taken to be a separable one,
making no use of quantum correlations \cite{MatsumotoPure}.
(See also Section B.3 in \cite{YFGsupp} for a simple proof.)
Note that the matrix $J^{(R)^{-1}}$ is degenerate, and the derived
quantum Gaussian shift model $ \{ N(h,J^{(R)^{-1}}) \}_h $ is
a canonical coherent model \cite{fujiwaraCoherent}.
\end{example}

%
\begin{example}[(2-dimensional faithful state model)]
The third example treats the case when the SLD tangent space is not $\D
$ invariant.
Let us consider the model
\[
\S= \bigl\{ \rho_{\theta} = \tfrac{1}{2} \bigl( I+
\theta^{1}\sigma_{1}+\theta^{2}
\sigma_{2}+ z_0 \sqrt{1-\|\theta\|^2}
\sigma_{3} \bigr); \theta= \bigl(\theta^i
\bigr)_{1\le i\le2}\in\Theta \bigr\},
\]
where $0 \leq z_0 <1$, and $\Theta$ is the open unit disk.
Due to the rotational symmetry around $z$-axis, we take the reference
point to be $\theta_0=(0,r)$, with $0\le r<1$.
By a direct calculation, we see that the SLDs at $\theta_0$ are
$(L_{1}, L_2)
= (
\sigma_{1},
\frac{1}{1-r^2}(\sigma_2-r I)
)$.
It is important to notice that the SLD tangent space $\span\{
L_{i} \} _{i=1}^{2}$ is not $\D$ invariant unless $r=0$. In fact
\[
\D\sigma_1=z(r) \sigma_2-r\sigma_3,\qquad \D
\sigma_2=-z(r) \sigma_1,
\]
where $z(r):=E[\sigma_3]=z_0\sqrt{1-r^2}$.
The minimal $\D$ invariant extension $\T$ of the SLD tangent space has
a basis $(D_1, D_2, D_3):=(L_{1}, L_{2}, \sigma_{3}-z(r) I)$.
The matrices~$\Sigma$, $J$, and $\tau$ appeared in Definition~\ref
{defQLANX} and Corollary~\ref{corqlecam3iid} are calculated as
\begin{eqnarray*}
\Sigma&:=& [\Tr\rho_{\theta_{0}}D_{j}D_{i}
]_{ij}\\
& =&\pmatrix{ 1 & -\displaystyle\i\frac{z_0^2}{z(r)} & r \i-z(r)
\vspace*{2pt}\cr
\i
\displaystyle\frac{z_0^2}{z(r)} & \displaystyle\frac{z_0^2}{z(r)^2} & - \displaystyle\biggl(\frac{r}{z(r)} + \i
\biggr)z_0^2
\vspace*{2pt}\cr
-r \i-z(r) & - \displaystyle\biggl(\frac{r}{z(r)}
- \i \biggr)z_0^2 & 1 },
\\
J&:=& [\Tr\rho_{\theta_{0}}L_{j}L_{i} ]_{ij} =
\pmatrix{ 1 & -\displaystyle\i\frac{z_0^2}{z(r)}
\vspace*{2pt}\cr
\displaystyle\i\frac{z_0^2}{z(r)} & \displaystyle\frac{z_0^2}{z(r)^2}
},
\\
\tau&:=& [\Tr\rho_{\theta_{0}}L_{j}\sigma_{i}
]_{ij}= \pmatrix{ 1 & -\displaystyle\i\frac{z_0^2}{z(r)}
\vspace*{2pt}\cr
\displaystyle\i\frac{z_0^2}{z(r)} &
\displaystyle\frac{z_0^2}{z(r)^2}
\vspace*{2pt}\cr
-r \i-z(r) & - \displaystyle\biggl(\frac{r}{z(r)} - \i
\biggr)z_0^2 }.
\end{eqnarray*}

Given a $2\times2$ real positive definite matrix $G$, the minimal
value of the weighted covariances at $\theta=\theta_0$ is given by
\[
\min_{\hat{M}}\Tr GV_{\theta_{0}}[\hat{M}]=C_{\theta_{0}}^{(1)}
(\rho_{\theta},G ),
\]
where the minimum is taken over all estimators $\hat M$ that are
locally unbiased at $\theta_0$, and
\[
C_{\theta_{0}}^{(1)} (\rho_{\theta},G ) = \bigl( \Tr\sqrt{
\sqrt{G} J^{(S)^{-1}} \sqrt{G} } \bigr)^{2}
\]
is the Nagaoka bound \cite{Nagaoka} (see also \cite{YamagataTomo}).

It can be shown that the Nagaoka bound is greater than the Holevo bound:
\[
C_{\theta_0}^{(1)} (\rho_{\theta},G )>C_{\theta_{0}} (\rho
_{\theta},G ).
\]
Let us check this fact for the special case when $G=J^{(S)}$.
A direct computation shows that
\[
C_{\theta_0}^{(1)} \bigl(\rho_{\theta},J^{(S)}
\bigr)=4
\]
and
\begin{eqnarray*}
C_{\theta_0} \bigl(\rho_{\theta},J^{(S)} \bigr) & = & \cases{
2(1+z_0)-r^2 \bigl(1-z_0^2
\bigr), & \quad $\mbox{if } \displaystyle 0 \leq r
\leq\sqrt{\frac{z_0}{1-z_0^2}}$,
\vspace*{2pt}\cr
2 +
\displaystyle\frac{z_0^2}{r^2 (1-z_0^2)}, & \quad $\mbox{if } \displaystyle\sqrt{\frac
{z_0}{1-z_0^2}}<r$. }
\end{eqnarray*}
The right panel of Figure~\ref{figbounds} shows the behaviour of
$C_{\theta_0} (\rho_{\theta},J^{(S)} )$ (solid) and $C_{\theta
_0}^{(1)} (\rho_{\theta},J^{(S)} )$ with $z_0=\frac{1}{4}$
(dashed) as functions of $r$.
We see that Holevo bound $C_{\theta_0} (\rho_{\theta},J^{(S)} )$
is much smaller than $C_{(0,r)}^{(1)} (\rho_{\theta},J^{(S)} )$.

As in Example~\ref{ex3d}, we demonstrate that the Holevo bound is
asymptotically achievable.
Let
\[
\Delta_{i}^{(n)}:=\frac{1}{\sqrt{n}}\sum
_{k=1}^{n}I^{\otimes
k-1}\otimes L_{i}
\otimes I^{\otimes n-k}\qquad (i=1,2),
\]
and let
\[
X_{j}^{(n)}:=\frac{1}{\sqrt{n}}\sum
_{k=1}^{n}I^{\otimes k-1}\otimes D_{j}
\otimes I^{\otimes n-k}\qquad (j=1,2,3).
\]
It then follows from the quantum central limit theorem that
\[
\biggl(\pmatrix{X^{(n)}
\cr
\Delta^{(n)} },
\rho_{\theta_{0}}^{\otimes n} \biggr)\mathop{\rightsquigarrow}_{q}N
\biggl(0,\pmatrix {\Sigma& \tau
\cr
\tau^{*} & J } \biggr).
\]
Therefore, Corollary~\ref{corqlecam3iid} shows that $ ( \{\rho
_\theta^{\otimes n} \},X^{(n)} )$ is jointly QLAN at $\theta
_0$, and that
\[
\bigl(X^{(n)},\rho_{\theta_{0}+h/\sqrt{n}}^{\otimes n} \bigr)\mathop{
\rightsquigarrow}_{q}N \bigl(( \Re\tau)h,\Sigma \bigr)
\]
for all $h\in\R^{2}$.

It should be noted that the off-diagonal block $\tau$ of the ``quantum
covariance'' matrix is not a square matrix.
This means that the derived quantum Gaussian shift model $ \{ N((\Re
\tau)h,\Sigma); h\in\R^{2} \} $ forms a submanifold of the total
quantum Gaussian shift model derived in Example~\ref{ex3d},
corresponding to a 2-dimensional linear subspace in the shift parameter space.
Nevertheless, Theorem~\ref{thmachieveHolevo} asserts that
there exists a sequence $M^{(n)}$ of estimators on the model
$ \{ \rho_{\theta_{0}+h/\sqrt{n}}^{\otimes n}; h\in\R^{3} \} $
that is asymptotically unbiased and achieves the Holevo bound:
\[
\lim_{n \rightarrow\infty} \Tr G V_{h}^{(n)}
\bigl[M^{(n)} \bigr] =C_{h} \bigl(N \bigl((\Re\tau)h,\Sigma
\bigr),G \bigr) =C_{\theta_0} (\rho_{\theta}, G )
\]
for all $h$ that belong to a dense subset of $\R^3$.
\end{example}

\subsection{\texorpdfstring{Translating estimation of $h$ to estimation of $\theta$}{Translating estimation of $h$ to estimation of theta}}

As we have seen in the previous subsections, our theory enables us to
construct asymptotically optimal estimators of $h$ in the local
models\vadjust{\goodbreak}
indexed by the parameter $\theta_0 + h/\sqrt{n}$. In practice of
course, $\theta_0$ is unknown and hence estimation of $h$, with
$\theta
_0$ known, is irrelevant. The actual sequence of measurements which we
have constructed depends in all interesting cases on $\theta_0$.

However, the results immediately inspire two-step (or adaptive)
procedures, in which we first measure a small proportion of the quantum
systems, in number $n_1$ say, using some standard measurement scheme,
for instance, separate particle quantum tomography. From these
measurement outcomes we construct an initial estimate of $\theta$, let
us call it $\widetilde\theta$. We can now use our theory to compute
the asymptotically optimal measurement scheme which corresponds to the
situation $\theta_0=\widetilde\theta$. We proceed to implement this
measurement on the remaining quantum systems collectively, estimating
$h$ in the model $\theta=\widetilde\theta+ h/\sqrt{n_2}$ where $n_2$
is the number of systems still available for the second stage.

What can we say about such a procedure? If $n_1/n\to\alpha>0$ as
$n\to\infty$, then we can expect that the initial estimate
$\widetilde
\theta$ is root $n$ consistent. In smooth models, one would expect that
in this case the final estimate $\widehat\theta= \widetilde\theta+
\widehat h/\sqrt{n_2}$ would be asymptotically optimal \emph{up to a
factor $1-\alpha$}: its limiting variance will be a factor $(1-\alpha
)^{-1}$ too large.

If however $n_1\to\infty$ but $n_1/n\to\alpha=0$, then one would
expect this procedure to break down, unless the rate of growth of $n_1$
is very carefully chosen (and fast enough). On the other hand, instead
of a direct two-step procedure, with the final estimate computed as
$\widetilde\theta+ \widehat h/\sqrt{n_2}$, one could be more careful
in how the data obtained from the second stage measurement is used.
Given the second step measurement, which results in an observed value
$\widehat h$, one could write down the likelihood for $h$ based on the
given measurement and the initially specified model, and compute
instead of the just mentioned one-step iterate, the actual maximum
likelihood estimator of $\theta$ based on the second stage data. Such
procedures have earlier been studied by Gill and Massar \cite
{GillMassar} and others, and shown in special cases to perform very well.

However, in general, the computational problem of even calculating the
likelihood given data, measurement, and model, is challenging, due to
the huge size of the Hilbert space of $n$ copies of a finite
dimensional quantum system.

\section{Concluding remarks}

We have developed a new theory of local asymptotic normality in the
quantum domain based on a quantum extension of the log-likelihood ratio.
This formulation is applicable to any model satisfying a mild
smoothness condition, and is free from artificial setups such as the
use of a special coordinate system and/or nondegeneracy of eigenvalues
of the reference state.
We also have proved asymptotic achievability of the Holevo bound for
the local shift parameter on a dense subset of the parameter space.

There are of course many open questions left.
Among others, it is not clear whether every sequence of statistics on a
QLAN model can be realised on the limiting quantum Gaussian shift model.\vadjust{\goodbreak}
In classical statistics, such a problem has been solved affirmatively
as the representation theorem, which asserts that, given a weakly
convergent sequence $T^{(n)}$ of statistics on $ \{ p_{\theta
_{0}+h/\sqrt{n}}^{(n)}; h\in\R^{d} \}$, there exist a limiting
statistics $T$ on\vspace*{-1pt} $ \{N(h,J^{-1}); h\in\R^{d} \}$ such that
$T^{(n)}\stackrel{h}{\rightsquigarrow} T$.
Representation theorem is useful in proving, for example, the
nonexistence of an asymptotically superefficient estimator (the
converse part, as stated in \hyperref[sec1]{Introduction}).
Moreover, the so-called convolution theorem and local asymptotic
minimax theorem, which are the standard tools in discussing asymptotic
lower bounds for estimation in LAN models, immediately follows \cite{Vaart}.
Extending the representation theorem, convolution theorem, and local
asymptotic minimax theorem to the quantum domain is an intriguing open
problem. However it surely is possible to make some progress in this
direction, as, for instance, the results of Gill and Gu\c{t}\u{a} \cite
{GillGuta}.
In that paper, the van Trees inequality was used to derive some results
in a ``poor man's'' version of QLAN theory;
see also \cite{vanTrees}.

It also remains to be seen whether our asymptotically optimal
statistical procedures for the local model with local parameter $h$ can
be translated into useful statistical procedures for the real world
case in which $\theta_0$ is unknown.


\begin{supplement}[id=suppA]
\stitle{Supplementary material to ``Quantum local asymptotic normality
based on a new quantum likelihood ratio''}
\slink[doi]{10.1214/13-AOS1147SUPP} 
\sdatatype{.pdf}
\sfilename{aos1147\_supp.pdf}
\sdescription{Section A is devoted to proofs of Lemma~\ref{lemoclt}, Theorems~\ref
{thmqlecam3} and \ref{thmQLANiid}, Corollary~\ref{corqlecam3iid},
and Theorem~\ref{thmachieveHolevo}.
Section B is devoted to a brief account of quantum estimation theory,
including quantum logarithmic derivatives, the commutation operator,
the Holevo bound,
estimation theory for quantum Gaussian shift models and for pure state models.}
\end{supplement}

%

\printaddresses


\begin{thebibliography}{24}

\bibitem{AmariNagaoka}
%
\begin{bbook}[mr]
\bauthor{\bsnm{Amari},~\bfnm{Shun-ichi}\binits{S.-i.}} \AND
\bauthor{\bsnm{Nagaoka},~\bfnm{Hiroshi}\binits{H.}}
(\byear{2000}).
\btitle{Methods of Information Geometry}.
\bseries{Translations of Mathematical Monographs}
\bvolume{191}.
\bpublisher{Amer. Math. Soc.}, \blocation{Providence, RI}.
\bid{mr={1800071}}
\bptok{imsref}%
\end{bbook}
%
\endbibitem

\bibitem{FujiwaraNagaoka1995}
%
\begin{barticle}[mr]
\bauthor{\bsnm{Fujiwara},~\bfnm{Akio}\binits{A.}} \AND
\bauthor{\bsnm{Nagaoka},~\bfnm{Hiroshi}\binits{H.}}
(\byear{1995}).
\btitle{Quantum {F}isher metric and estimation for pure state models}.
\bjournal{Phys. Lett. A}
\bvolume{201}
\bpages{119--124}.
\bid{doi={10.1016/0375-9601(95)00269-9}, issn={0375-9601}, mr={1329961}}
\bptok{imsref}%
\end{barticle}
%
\endbibitem

\bibitem{fujiwaraCoherent}
%
\begin{barticle}[mr]
\bauthor{\bsnm{Fujiwara},~\bfnm{Akio}\binits{A.}} \AND
\bauthor{\bsnm{Nagaoka},~\bfnm{Hiroshi}\binits{H.}}
(\byear{1999}).
\btitle{An estimation theoretical characterization of coherent states}.
\bjournal{J. Math. Phys.}
\bvolume{40}
\bpages{4227--4239}.
\bid{doi={10.1063/1.532962}, issn={0022-2488}, mr={1708381}}
\bptok{imsref}%
\end{barticle}
%
\endbibitem

\bibitem{GillGuta}
%
\begin{barticle}[auto:STB|2013/06/05|13:45:01]
\bauthor{\bsnm{Gill},~\bfnm{R.~D.}\binits{R.~D.}} \AND
\bauthor{\bsnm{Gu{\c{t}}{\u{a}}},~\bfnm{M.}\binits{M.}}
(\byear{2012}).
\btitle{On asymptotic quantum statistical inference}.
\bjournal{IMS Collections From Probability to Statistics and Back:
High-Dimensional Models and Processes}
\bvolume{9}
\bpages{105--127}.
\bptok{imsref}%
\end{barticle}
%
\endbibitem

\bibitem{vanTrees}
%
\begin{barticle}[mr]
\bauthor{\bsnm{Gill},~\bfnm{Richard~D.}\binits{R.~D.}} \AND
\bauthor{\bsnm{Levit},~\bfnm{Boris~Y.}\binits{B.~Y.}}
(\byear{1995}).
\btitle{Applications of the {V}an {T}rees inequality: A~{B}ayesian
{C}ram\'er--{R}ao bound}.
\bjournal{Bernoulli}
\bvolume{1}
\bpages{59--79}.
\bid{doi={10.2307/3318681}, issn={1350-7265}, mr={1354456}}
\bptok{imsref}%
\end{barticle}
%
\endbibitem

\bibitem{GillMassar}
%
\begin{barticle}[auto:STB|2013/06/05|13:45:01]
\bauthor{\bsnm{Gill},~\bfnm{R.~D.}\binits{R.~D.}} \AND
\bauthor{\bsnm{Massar},~\bfnm{S.}\binits{S.}}
(\byear{2000}).
\btitle{State estimation for large ensembles}.
\bjournal{Phys. Rev. A (3)}
\bvolume{61}
\bpages{042312}.
\bptok{imsref}%
\end{barticle}
%
\endbibitem

\bibitem{GutaUsta}
%
\begin{barticle}[mr]
\bauthor{\bsnm{Gu{\c{t}}{\u{a}}},~\bfnm{M{\u{a}}d{\u
{a}}lin}\binits
{M.}} \AND
\bauthor{\bsnm{Butucea},~\bfnm{Cristina}\binits{C.}}
(\byear{2010}).
\btitle{Quantum {$U$}-statistics}.
\bjournal{J. Math. Phys.}
\bvolume{51}
\bpages{102202, 24}.
\bid{doi={10.1063/1.3476776}, issn={0022-2488}, mr={2761295}}
\bptok{imsref}%
\end{barticle}
%
\endbibitem

\bibitem{GutaQLANweak}
%
\begin{barticle}[mr]
\bauthor{\bsnm{Gu{\c{t}}{\u{a}}},~\bfnm{M{\u{a}}d{\u
{a}}lin}\binits
{M.}} \AND
\bauthor{\bsnm{Jen{\v{c}}ov{\'a}},~\bfnm{Anna}\binits{A.}}
(\byear{2007}).
\btitle{Local asymptotic normality in quantum statistics}.
\bjournal{Comm. Math. Phys.}
\bvolume{276}
\bpages{341--379}.
\bid{doi={10.1007/s00220-007-0340-1}, issn={0010-3616}, mr={2346393}}
\bptok{imsref}%
\end{barticle}
%
\endbibitem

\bibitem{GutaQLANfor2}
%
\begin{barticle}[mr]
\bauthor{\bsnm{Gu{\c{t}}{\u{a}}},~\bfnm{M{\u{a}}d{\u
{a}}lin}\binits
{M.}} \AND
\bauthor{\bsnm{Kahn},~\bfnm{Jonas}\binits{J.}}
(\byear{2006}).
\btitle{Local asymptotic normality for qubit states}.
\bjournal{Phys. Rev. A (3)}
\bvolume{73}
\bpages{052108, 15}.
\bid{doi={10.1103/PhysRevA.73.052108}, issn={1050-2947}, mr={2229156}}
\bptok{imsref}%
\end{barticle}
%
\endbibitem

\bibitem{Hayashi}
%
\begin{bincollection}[auto:STB|2013/06/05|13:45:01]
\bauthor{\bsnm{Hayashi},~\bfnm{M.}\binits{M.}}
(\byear{1997}).
\btitle{A linear programming approach to attainable Cram\'er--Rao type bounds}.
In \bbooktitle{Quantum Communication, Computing, and Measurement}
\bpages{99--108}.
\bpublisher{Plenum}, \baddress{New York}.
\bptok{imsref}%
\end{bincollection}
%
\endbibitem

\bibitem{HayashiMatsumoto}
%
\begin{barticle}[mr]
\bauthor{\bsnm{Hayashi},~\bfnm{Masahito}\binits{M.}} \AND
\bauthor{\bsnm{Matsumoto},~\bfnm{Keiji}\binits{K.}}
(\byear{2008}).
\btitle{Asymptotic performance of optimal state estimation in qubit system}.
\bjournal{J. Math. Phys.}
\bvolume{49}
\bpages{102101, 33}.
\bid{doi={10.1063/1.2988130}, issn={0022-2488}, mr={2464597}}
\bptok{imsref}%
\end{barticle}
%
\endbibitem

\bibitem{Holevo}
%
\begin{bbook}[mr]
\bauthor{\bsnm{Holevo},~\bfnm{Alexander}\binits{A.}}
(\byear{2011}).
\btitle{Probabilistic and Statistical Aspects of Quantum Theory},
\bedition{2nd} ed.
\bseries{Quaderni. Monographs}
\bvolume{1}.
\bpublisher{Edizioni della Normale}, \blocation{Pisa}.
\bid{doi={10.1007/978-88-7642-378-9}, mr={2797301}}
\bptok{imsref}%
\end{bbook}
%
\endbibitem

\bibitem{qclt}
%
\begin{barticle}[mr]
\bauthor{\bsnm{Jak{\v{s}}i{\'c}},~\bfnm{V.}\binits{V.}},
\bauthor{\bsnm{Pautrat},~\bfnm{Y.}\binits{Y.}} \AND
\bauthor{\bsnm{Pillet},~\bfnm{C.~A.}\binits{C.~A.}}
(\byear{2010}).
\btitle{A quantum central limit theorem for sums of independent identically
distributed random variables}.
\bjournal{J. Math. Phys.}
\bvolume{51}
\bpages{015208, 8}.
\bid{doi={10.1063/1.3285287}, issn={0022-2488}, mr={2605841}}
\bptok{imsref}%
\end{barticle}
%
\endbibitem

\bibitem{GutaQLANforD}
%
\begin{barticle}[mr]
\bauthor{\bsnm{Kahn},~\bfnm{Jonas}\binits{J.}} \AND
\bauthor{\bsnm{Gu{\c{t}}{\u{a}}},~\bfnm{M{\u{a}}d{\u
{a}}lin}\binits{M.}}
(\byear{2009}).
\btitle{Local asymptotic normality for finite dimensional quantum systems}.
\bjournal{Comm. Math. Phys.}
\bvolume{289}
\bpages{597--652}.
\bid{doi={10.1007/s00220-009-0787-3}, issn={0010-3616}, mr={2506764}}
\bptok{imsref}%
\end{barticle}
%
\endbibitem

\bibitem{KuboAndo}
%
\begin{barticle}[mr]
\bauthor{\bsnm{Kubo},~\bfnm{Fumio}\binits{F.}} \AND
\bauthor{\bsnm{Ando},~\bfnm{Tsuyoshi}\binits{T.}}
(\byear{1979/80}).
\btitle{Means of positive linear operators}.
\bjournal{Math. Ann.}
\bvolume{246}
\bpages{205--224}.
\bid{doi={10.1007/BF01371042}, issn={0025-5831}, mr={0563399}}
\bptnote{check year}%
\bptok{imsref}%
\end{barticle}
%
\endbibitem

\bibitem{CCR1}
%
\begin{barticle}[mr]
\bauthor{\bsnm{Manuceau},~\bfnm{J.}\binits{J.}},
\bauthor{\bsnm{Sirugue},~\bfnm{M.}\binits{M.}},
\bauthor{\bsnm{Testard},~\bfnm{D.}\binits{D.}} \AND
\bauthor{\bsnm{Verbeure},~\bfnm{A.}\binits{A.}}
(\byear{1973}).
\btitle{The smallest {$C^{\ast} $}-algebra for canonical commutations
relations}.
\bjournal{Comm. Math. Phys.}
\bvolume{32}
\bpages{231--243}.
\bid{issn={0010-3616}, mr={0339715}}
\bptok{imsref}%
\end{barticle}
%
\endbibitem

\bibitem{MatsumotoPure}
%
\begin{barticle}[mr]
\bauthor{\bsnm{Matsumoto},~\bfnm{K.}\binits{K.}}
(\byear{2002}).
\btitle{A new approach to the {C}ram\'er--{R}ao-type bound of the pure-state
model}.
\bjournal{J. Phys. A}
\bvolume{35}
\bpages{3111--3123}.
\bid{doi={10.1088/0305-4470/35/13/307}, issn={0305-4470}, mr={1913859}}
\bptok{imsref}%
\end{barticle}
%
\endbibitem

\bibitem{Nagaoka}
%
\begin{barticle}[auto:STB|2013/06/05|13:45:01]
\bauthor{\bsnm{Nagaoka},~\bfnm{H.}\binits{H.}}
(\byear{1991}).
\btitle{A generalization of the simultaneous diagonalization of Hermitian
matrices and its relation to quantum estimation theory (in Japanese)}.
\bjournal{Transactions of the Japan Society for Industrial and Applied
Mathematics}
\bvolume{1}
\bpages{305--318}.
\bptok{imsref}%
\end{barticle}
%
\endbibitem

\bibitem{CCR2}
%
\begin{bbook}[mr]
\bauthor{\bsnm{Petz},~\bfnm{D{\'e}nes}\binits{D.}}
(\byear{1990}).
\btitle{An Invitation to the Algebra of Canonical Commutation Relations}.
\bseries{Leuven Notes in Mathematical and Theoretical Physics. Series A:
Mathematical Physics}
\bvolume{2}.
\bpublisher{Leuven Univ. Press}, \blocation{Leuven}.
\bid{mr={1057180}}
\bptok{imsref}%
\end{bbook}
%
\endbibitem

\bibitem{Petz}
%
\begin{bbook}[mr]
\bauthor{\bsnm{Petz},~\bfnm{D{\'e}nes}\binits{D.}}
(\byear{2008}).
\btitle{Quantum Information Theory and Quantum Statistics}.
\bpublisher{Springer}, \blocation{Berlin}.
\bid{mr={2363070}}
\bptnote{check year}%
\bptok{imsref}%
\end{bbook}
%
\endbibitem

\bibitem{Vaart}
%
\begin{bbook}[mr]
\bauthor{\bparticle{van~der} \bsnm{Vaart},~\bfnm{A.~W.}\binits{A.~W.}}
(\byear{1998}).
\btitle{Asymptotic Statistics}.
\bseries{Cambridge Series in Statistical and Probabilistic Mathematics}
\bvolume{3}.
\bpublisher{Cambridge Univ. Press}, \blocation{Cambridge}.
\bid{mr={1652247}}
\bptok{imsref}%
\end{bbook}
%
\endbibitem

\bibitem{YamagataTomo}
%
\begin{barticle}[mr]
\bauthor{\bsnm{Yamagata},~\bfnm{Koichi}\binits{K.}}
(\byear{2011}).
\btitle{Efficiency of quantum state tomography for qubits}.
\bjournal{Int. J. Quantum Inf.}
\bvolume{9}
\bpages{1167--1183}.
\bid{doi={10.1142/S0219749911007551}, issn={0219-7499}, mr={2931457}}
\bptok{imsref}%
\end{barticle}
%
\endbibitem

\bibitem{YFGsupp}
%
\begin{bmisc}[auto:STB|2013/06/05|13:45:01]
\bauthor{\bsnm{Yamagata},~\bfnm{K.}\binits{K.}},
\bauthor{\bsnm{Fujiwara},~\bfnm{A.}\binits{A.}} \AND
\bauthor{\bsnm{Gill},~\bfnm{R.~D.}\binits{R.~D.}}
(\byear{2013}).
\bhowpublished{Supplement to ``Quantum local asymptotic
normality based
on a new quantum likelihood ratio.'' DOI:\doiurl{10.1214/13-AOS1147SUPP}.}
\bptok{imsref}%
\end{bmisc}
%
\endbibitem

\end{thebibliography}
\end{document}